\documentclass[10pt,aps,pra, twocolumn, superscriptaddress,floatfix,nofootinbib]{revtex4-1}
\usepackage{graphicx}
\usepackage{subfigure}
\usepackage[english]{babel}
\usepackage{amsmath}
\usepackage{amssymb}

\usepackage[usenames,dvipsnames]{color}
\usepackage{color}
\definecolor{Blue}{rgb}{0.3,0.3,0.9}
\definecolor{orange}{rgb}{1,0.5,0}

\newcommand{\bI}{\mathbb I}
\newcommand{\bS}{S}
\newcommand{\bbS}{\mathbb S}
\newcommand{\bU}{U}
\newcommand{\bbU}{\mathbb U}
\newcommand{\inn}{\mbox{\scriptsize  in}}
\newcommand{\out}{\mbox{\scriptsize  out}}

\newcommand{\dd}{\mbox{d}}
\newcommand{\tr}{\mbox{\mbox{Tr}}}
\newcommand{\dS}{\mbox{d}{\mathbf\Sigma}}
\newcommand{\bx}{{\mathbf x}}

\newcommand{\bk}{{\mathbf k}}
\newcommand{\bq}{{\mathbf K}}
\newcommand{\bv}{{\mathbf v}}
\newcommand{\sgn}{\mbox{sgn}}
\newcommand{\im}{\,{ \rm Im}\, }
\newcommand{\re}{\,{ \rm Re}\, }
\newcommand{\pr}{\,{ \rm P}\, }

\begin{document}

\title{A Scattering Approach to the Dynamical Casimir Effect}
\author{Mohammad F. Maghrebi}
\affiliation{Center for Theoretical Physics, Massachusetts Institute of Technology, Cambridge, MA 02139, USA}
\affiliation{Department of Physics, Massachusetts Institute of Technology, Cambridge, MA 02139, USA}
\author{Ramin Golestanian}
\affiliation{Rudolf Peierls Centre for Theoretical Physics, University of Oxford, Oxford OX1 3NP, UK}
\author{Mehran Kardar}
\affiliation{Department of Physics, Massachusetts Institute of Technology, Cambridge, MA 02139, USA}

\begin{abstract}
  We develop a unified scattering approach to dynamical Casimir problems which can be applied to both accelerating boundaries, as well as dispersive objects in relative motion. A general (trace) formula is derived for the radiation from accelerating boundaries. Applications are provided for objects with different shapes in various dimensions, and undergoing rotational or linear motion. Within this framework, photon generation is discussed in the context of a modulated optical mirror. For dispersive objects, we find general results solely in terms of the scattering matrix. Specifically, we discuss the vacuum friction on a rotating object, and the friction on an \emph{atom }moving parallel to a surface.

\end{abstract}

\maketitle

\section{Introduction}
Quantum zero-point fluctuations manifest themselves in a variety of macroscopic effects. A prominent example is Casimir's demonstration that these fluctuations lead to attraction of two perfectly conducting parallel plates \cite{Casimir48-2}. Experimental advances in precision measurements of the Casimir force \cite{Lamoreaux97, Mohideen98} have revived interest in finding frameworks where one can compute these forces both numerically \cite{Reid09,Rodriguez10} and analytically. A particularly successful approach in applications to different geometries and material properties is based on scattering methods and techniques \cite{Emig07,Kenneth08,Maia_Neto08,Emig08-1,Rahi09, Canaguier-Durand10, Maghrebi10}. In this approach, the quantum-field-theoretic problem is reduced to that of finding the {\it classical} scattering matrix of each object.

Another manifestation of fluctuations appears in the so-called dynamical Casimir effect: when objects are set in motion, they interact with the fluctuations of the background vacuum in a time-dependent fashion which excites photons and emits radiation. In fact, accelerating boundaries radiate energy and thus experience friction. An early example of this phenomenon was discussed by Moore for a one dimensional cavity \cite{Moore70}. A relativistic analysis of an accelerating mirror in 1+1 dimension in Ref.~\cite{Fulling76} employs techniques from  conformal field theory. A perturbative study of the latter confirmed and generalized its results to higher dimensions \cite{Ford82}. Among other methods, the fluctuation-dissipation theorem has been used to compute the frictional force on a moving sphere in free space \cite{MaiaNeto93}, a Hamiltonian formalism has been applied to the problem of photon production in cavities \cite{Law94,Dodonov95}, and a (Euclidean) path-integral formulation is introduced to study the ``vacuum'' friction for a rough plate moving laterally \cite{Golestanian97}. We specially note that an input-output formalism relating the incoming and outgoing operators is used to compute, among other things, the frequency and angular spectrum of radiated photons \cite{Neto96}. While a substantial literature is devoted to objects with perfect boundary conditions, dielectric and dispersive materials have also been studied in some cases \cite{Barton93}. In fact, dispersive objects exhibit similar effects even when they move at a constant velocity. For example, two parallel plates moving laterally with respect to each other experience a (non-contact) frictional force \cite{Pendry97,Volokitin99}. Even a single object experiences friction if put in constant rotation \cite{Manjavacas10, Maghrebi12}, a phenomenon most intimately related to \emph{superradiance }first discovered by Zel'dovich \cite{Zel'dovich71}. (Translational motion of a single object is trivial due to the Lorentz symmetry.) The latter examples, consisting of dispersive objects moving at a constant rate, are usually treated within the framework of the fluctuation-dissipation theorem or the closely related Rytov formalism \cite{Rytov89}.

Inspection of the literature on the dynamical Casimir effect leads to the following observations: There are a plethora of interesting---sometimes counter-intuitive--- phenomena emerging from the motion of a body in an ambient quantum field \cite{Dodonov10,Dalvit11}. These phenomena span a number of subfields in physics, and have been treated by a variety of different formalisms. Even the simplest examples appear to require rather complex computations. Only recently experimental realizations---using a SQUID to mimic the moving boundary of a cavity (transmission line) \cite{Nori11,Nori12}---have made precise measurements possible, raising the hope for an explosion of activity similar to the post-precision experiment era of static Casimir forces. This motivates reexamination of theoretical literature on the subject, aiming for a simple and unifying framework for analysis.

In this work, we follow two goals. First, inspired by the success of the scattering-theory methods in (static) Casimir forces, we attempt at extending these techniques to dynamical Casimir problems. We find that the \emph{classical} scattering matrix is naturally incorporated into the formalism. However, dynamical configurations provide new channels where the incoming frequency jumps to different values, hence the scattering matrix should be defined accordingly.
Second, we aim for a universal framework which brings the diverse set of problems in dynamical Casimir under the same rubric. Most notably, we treat accelerating boundaries, modulated optical setups and moving dispersive objects---usually tackled with different techniques, as explained above---on the same footing.

In this paper, computations are performed for a scalar field theory. The generalization to electromagnetism is straightforward in principle, while practical computations are more complicated in the latter. We find scalar field theory convenient to set the framework for more realistic applications.

The paper is organized as follows: In Sec. \ref{Sec: Formalism}, starting from a second-quantized formalism, we derive general formulas for the energy radiation due to the dynamical Casimir effect. In Sec. \ref{Sec: Lossless accelerating objects}, we consider lossless objects undergoing non-uniform motion or optical modulation, and provide a variety of examples to showcase the power of the scattering approach. Specifically, we find that an (asymmetrical) spinning object slows down, and further contrast linear and angular motion. In Sec. \ref{Sec. Lossy objects: stationary motion}, we consider dispersive objects moving at a constant rate. We also generalize to the case of multiple objects in relative motion where we study, among other things, an ``atom'' moving parallel to a dispersive surface.

\section{Formalism}\label{Sec: Formalism}

 We start with input-output relations as described in Ref.~\cite{Beenakker98}. The underlying formalism has been developed to quantize the electromagnetic field in a lossy or amplifying medium \cite{Matloob95,Gruner96, Loudon97}. (A similar method is also used to study the dynamical Casimir effect; see, for example, Refs.~\cite{Lambrecht96, Neto96}. However the more general formalism in Ref.~\cite{Beenakker98} allows further extensions specially to dispersive objects.) Within this formalism the operators $\hat a^{\inn}$ and $\hat a^{\out}$ represent annihilation operators of the incoming or outgoing waves, respectively, in the vacuum (outside the object). These operators are then related by \cite{Beenakker98}
 \begin{equation}\label{Eq: 1}
    \hat a^{\out}_\beta= \sum_\alpha \bS_{\beta\alpha}\, \hat a_\alpha^{\inn} +\sum_\alpha\bU_{\beta\alpha} \, \hat b_{\alpha}\,,
 \end{equation}
 where $\hat b$ is the operator corresponding to the absorption within the object, and  $\alpha$ and $\beta$ are quantum numbers. In this equation, $\bS$ is the object's scattering matrix while $\bU$ describes its lossy character---the latter is related to the scattering matrix as shown later. This method treats field theory in the second quantized picture where quantum (annihilation or creation) operators are introduced. Equation (\ref{Eq: 1}) then relates quantum operators via the \emph{classical }scattering matrix. This proves to be useful in applications to the dynamical Casimir effect.

 Equation (\ref{Eq: 1}) has its roots in the classical wave equation. To see this, we first define ``in'' and ``out'' wave functions. The incoming wave can be expanded as
 \begin{equation}\label{Eq: incoming wave}
    \Phi^{\inn}(\bx)=\sum_\alpha c^{\inn}_\alpha \Phi^{\inn}_\alpha(\bx),
 \end{equation}
  where $c^{\inn}_\alpha$ is the amplitude of the corresponding wave function $\Phi^{\inn}_\alpha$. The latter function should be normalized so that the number of incoming quanta per unit time is (negative) unity,
  \begin{equation}\label{Eq: normalization1}
      \frac{1}{2i} \oint \dS \cdot \left[{\Phi^{\inn*}_{\alpha}} \, \nabla {\Phi^{\inn}_{\beta}}-\nabla {\Phi^{\inn*}_{\alpha}} \, {\Phi^{\inn}_{\beta}}\right]=-\delta_{\alpha \beta},
  \end{equation}
  with the integral defined over a closed surface enclosing the object.
  The reason for this choice is that we shall associate the wave function with a quantum operator which satisfies the canonical commutation relations, and the normalization should be defined consistently. Similarly the outgoing wave functions are normalized as
  \begin{equation}\label{Eq: normalization2}
      \frac{1}{2i} \oint \dS \cdot \left[{\Phi^{\out*}_{\alpha}} \, \nabla {\Phi^{\out}_{\beta}}-\nabla {\Phi^{\out*}_{\alpha}} \, {\Phi^{\out}_{\beta}}\right]=\delta_{\alpha \beta}.
  \end{equation}
    We also designate the solutions to the wave equation inside the object as $\Phi^{\rm obj}$. Now suppose that there are sources in two regions in space: at infinity where they generate the incoming wave, $\Phi^{\inn}$ in Eq.~(\ref{Eq: incoming wave}); and within the object where they induce a field $\Phi^{\rm obj}=\sum_\alpha d_\alpha \Phi_\alpha^{\rm obj}$---the normalization of these functions does not affect the scattering matrix and thus is not discussed here. The incoming wave is scattered by the object while the object itself radiates due to the induced field. The resulting outgoing wave, $\Phi^{\out}=\sum_\alpha c_\alpha^{\out} \Phi_\alpha^{\out}$, is determined by
  \begin{equation}\label{Eq: classical amplitude}
    c^{\out}_\beta=\sum_\alpha S_{\beta \alpha} c^{\inn}_\alpha + \sum_\alpha U_{\beta \alpha} d_\alpha,
  \end{equation}
  where the first term is merely the scattering of the incoming waves, and the second term captures the radiation of the object itself. The above analysis is based on a classical wave equation. Equation (\ref{Eq: 1}) extends the last equation to a relation between quantum operators, i.e. the complex-valued coefficients in Eq.~(\ref{Eq: classical amplitude}) become quantum operators in Eq.~(\ref{Eq: 1}) through $c^{\inn/\out} \to \hat a^{\inn/\out}$ and $d \to \hat b$ (see also the discussion in Ref.~\cite{Scully1997} on the relation between wave mechanics and the classical limit). From this point on, we shall drop the hat symbol from quantum operators.

  For objects in motion, scattering can also change the frequency of the incoming wave. We make the dependence on frequency explicit while reserving $\alpha$ for other quantum numbers; the sum in Eq.~(\ref{Eq: 1}) is replaced by
  \begin{equation}
     \sum_\alpha\int_{-\infty}^{\infty} \frac{\dd\omega}{2\pi}. \nonumber
  \end{equation}
  Most importantly, the scattering matrix may mix positive and negative frequencies, in which case an outgoing operator of positive frequency is related to an incoming operator of negative frequency via Eq.~(\ref{Eq: 1}). Note that an operator $a_{\omega \alpha}$ with negative $\omega$ should be interpreted as a creation operator; more precisely, $a_{\omega \alpha}=a^\dagger_{-\omega \,\bar \alpha}$ where $\bar \alpha$ is related to $\alpha$ by time reversal.

  We assume that the environment is at a temperature $T_{\rm env}$ while the object is at a (possibly different) temperature $T$. The distribution of the incoming modes (before scattering)  is solely characterized by $T_{\rm env}$,
  \begin{equation}\label{Eq: Occ Number}
    \langle {a^{\inn \dagger}_{\omega'\beta}} a^{\inn}_{\omega\alpha}\rangle= \sgn(\omega) n(\omega,T_{\rm env})\, \delta (\omega-\omega')\,\delta_{\alpha\beta}\,,
  \end{equation}
  where $n(\omega,T)=\frac{1}{\exp({\hbar \omega}/{k T})-1}$ is the Bose-Einstein factor. Note that this equation holds for both positive and negative values of frequency; an operator $a_{\omega \alpha}$ ($a^\dagger_{\omega \alpha}$) defined at a negative frequency, $\omega$, is interpreted as a creation (annihilation) operator of a positive-frequency mode.

  On the other hand, the occupation number for the operators $b$, localized on the object, is determined by $T$, the object's temperature. But we should keep in mind that the object could be moving, so the frequency defined from the point of view of a reference system (co)moving with the object is different from that of an observer in the vacuum, or the \emph{lab}, frame. For a partial wave $(\omega, \alpha)$ in the lab frame, we define $\tilde \omega_\alpha$ as the frequency according to the comoving reference frame. The occupation number is then\footnote{The change of basis from the frequency in the moving frame to that of the lab frame gives rise to the Jacobian. The partial derivative is positive on physical grounds.}
  \begin{equation}\label{Eq: Occ Number b}
    \langle {b}^\dagger_{\omega'\beta} b_{\omega\alpha}\rangle= \sgn(\tilde \omega_\alpha) n(\tilde \omega_\alpha,T)\, \frac{\partial \tilde\omega_\alpha}{\partial \omega}\, \delta(\omega-\omega')\,\delta_{\alpha\beta}\,.
  \end{equation}
  From the distribution of the incoming and localized operators, Eqs.~(\ref{Eq: Occ Number}) and (\ref{Eq: Occ Number b}), we can evaluate the distribution for an outgoing mode $(\omega', \beta)$,
  \begin{align}\label{Eq: Occ num out}
    \langle a^{\out \dagger}_{\omega'\beta} &a^{\out}_{\omega'\beta}\rangle= \int_{-\infty}^{\infty}\!\!\frac{\dd\omega}{2\pi} \, \sgn(\omega) n(\omega,T_{\rm env}) \sum_{\alpha}\left|\bS_{\omega'\beta,\omega\alpha}\right|^2 \nonumber \\
    &+ \sum_{\alpha}  \int_{-\infty}^{\infty}\!\!\frac{\dd\tilde \omega_\alpha}{2\pi} \,\sgn(\tilde \omega_\alpha) n(\tilde \omega_\alpha,T) \left|\bU_{\omega'\beta,\omega\alpha}\right|^2 .
  \end{align}
  To find the flux of field quanta to the environment, one should compute the difference of outgoing and incoming flux.
  To study Eq.~(\ref{Eq: Occ num out}) in some detail, we consider two different situations. \\

  {\bf Accelerating objects---}First we assume that the object is non-lossy so that the second term on the RHS of Eqs.~(\ref{Eq: 1}) and (\ref{Eq: Occ num out}) is absent. For the sake of simplicity, we choose to work at zero temperature\footnote{In the absence of loss, the object's temperature does not play a role.}, i.e. $T_{\rm env}=0$; the generalization to finite temperatures is straightforward. From Eq.~(\ref{Eq: Occ num out}), we find
  \begin{equation}\label{Eq: flux rate lossless}
    \langle a^{\out \dagger}_{\omega'\beta} a^{\out}_{\omega'\beta}\rangle= \int_{-\infty}^{0}\frac{\dd\omega}{2\pi} \sum_{\alpha}\left|\bS_{\omega'\beta,\omega\alpha}\right|^2.
  \end{equation}
  We have used the fact that the Bose-Einstein distribution at $T=0$ is different from zero only for negative frequencies. Loosely speaking, this means that, in the vacuum state, all single-particle states with negative energy are occupied while those of positive energy are empty.
  The rate of energy radiation is obtained as an integral over the outgoing flux in Eq.~(\ref{Eq: flux rate lossless}) multiplied by the quanta energy
  \begin{equation}\label{Eq: Enrgy Rad}
    P=\int_{0}^{\infty}\frac{\dd\omega'}{2\pi} \, \hbar \omega'     \int_{-\infty}^{0}\frac{\dd\omega}{2\pi} \sum_{\alpha,\beta}\left|\bS_{\omega'\beta,\omega\alpha}\right|^2.
  \end{equation}
  The choice of basis $\alpha$ is a matter of convenience, as in a basis-independent notation Eq.~(\ref{Eq: Enrgy Rad}) is cast as
  \begin{equation}\label{Eq: Enrgy Rad Trace}
    P=\int_{0}^{\infty}\frac{\dd\omega'}{2\pi} \, \hbar \omega'     \int_{-\infty}^{0}\frac{\dd\omega}{2\pi} \tr \left(\bbS_{\omega',\omega}\bbS^\dagger_{\omega',\omega}\right),
  \end{equation}
  where $\bbS$ is the basis-free scattering matrix. A similar expression is derived in Ref.~\cite{Lambrecht96} for the radiation from a vibrating cavity. These equations provide a simple and compact formulation which serve as the starting point for studying accelerating boundaries and modulated optical devices in Sec. \ref{Sec: Lossless accelerating objects}.\\

  {\bf Stationary motion---}Next we consider objects in stationary, linear or rotational, motion. Although the objects are moving, the boundaries do not change their shape or orientation. One such example is two infinite plates moving parallel to their surface. Despite the motion, the relative configuration of the two plates does not change in time.

  This type of dynamical problem is not explicitly time dependent, nevertheless the relative motion leads to dissipative effects. In other words, such systems respect time translation but break time-reversal symmetry, and thus allow for dissipation.  Because of the stationary character of the setup, however, the scattering matrix $\bbS$ as well as the matrix $\bbU$ are diagonal in frequency, and we indicate this by a single frequency dependence as $\bS_{\beta\alpha}(\omega)$ and $\bU_{\beta\alpha}(\omega)$.

  For a lossy object, the scattering matrix cannot be unitary as part of the incoming wave is lost inside the object. Interestingly, unitarity alone, sufficiently constrains the matrix $\bbU$ for our purposes \cite{Beenakker98}.
  There is a large body of literature on quantization in an absorbing (or amplifying) medium, covering a variety of approaches. The method of input-output relations \cite{Matloob95,Gruner96,Beenakker98} starts by formulating canonical commutation relations for the incoming and outgoing operators,
  \begin{align}
    &[a^{\out/ \inn}_{\omega\alpha}, {a^{\out/ \inn \dagger}_{\omega'\beta}}]=\sgn(\omega) \delta(\omega-\omega') \, \delta_{\alpha \beta}.
  \end{align}
  We extend this method to moving systems by demanding that the operators $b$ (localized on the object) satisfy the commutation relations in the rest frame of the object,
  \begin{align}
    &[b_{\omega\alpha},{b^\dagger_{\omega'\beta}}]=\sgn(\tilde \omega_\alpha)\frac{\partial \tilde \omega_\alpha}{\partial \omega }\delta(\omega -\omega')\, \delta_{\alpha\beta},
  \end{align}
  with $\tilde \omega_\alpha$ defined above. This set of relations along with Eq.~(\ref{Eq: 1}) lead to
  \begin{equation} \label{Eq: Unitarity moving}
   \sgn(\omega)\, (1-\sum_\alpha |\bS_{\beta \alpha}(\omega)|^2) =\sum_\alpha {\sgn(\tilde\omega_\alpha) }{\frac{\partial \tilde\omega_\alpha}{\partial \omega} }\, |\bU_{\beta\alpha}(\omega)|^2 \,.
  \end{equation}
  In the comoving frame, the object is momentarily at rest, and thus the frequency according to this frame does not change, i.e. $\tilde \omega_\alpha=\tilde \omega_\beta$ if $U_{\beta \alpha}\ne 0$. Therefore, Eq.~(\ref{Eq: Unitarity moving}) can be recast as
  \begin{equation} \label{Eq: Unitarity moving1}
    \sgn(\omega)\, (1-\sum_\alpha |\bS_{\beta \alpha}(\omega)|^2) ={\sgn(\tilde\omega_\beta) }{\frac{\partial \tilde\omega_\beta}{\partial \omega} }\sum_\alpha  |\bU_{\beta\alpha}(\omega)|^2 \,,
  \end{equation}
  or in a matrix notation,
  \begin{equation}\label{Eq: Unitarity moving2}
    \sgn(\omega)\, (\bI-\bbS\bbS^\dagger)_{\beta\beta} = {\sgn(\tilde\omega_\beta)}{\frac{\partial \tilde\omega_\beta}{\partial \omega} }(\bbU \bbU^\dagger)_{\beta \beta} \,,
  \end{equation}
  which constrains the matrix $\bbU$ in terms of the scattering matrix, $\bbS$. In
  the limit of static objects, one recovers the (basis-free) relation \cite{Beenakker98}
  \begin{equation}\label{Eq: Unitarity}
    \bI-\bbS \bbS^\dagger=\bbU \bbU^\dagger.
  \end{equation}
  This equation is interpreted by Beenakker as a ``fluctuation-dissipation'' relation with the LHS giving the dissipation due to the classical scattering from a lossy material, and the RHS accounting for  field fluctuations due to spontaneous absorption or emission (in an amplifying medium) of field quanta.

  Equation (\ref{Eq: Unitarity moving1}) for a moving object can be inserted in Eq.~(\ref{Eq: Occ num out}) to obtain the flux due to the outgoing quanta. We are interested in the total flux,
   \begin{equation}
   \frac{d\cal N}{\dd\omega}= \sum_{\beta} \, \langle a^{\out \dagger}_{\omega\beta} a^{\out}_{\omega\beta}\rangle-\langle a^{\inn \dagger}_{\omega\beta} a^{\inn}_{\omega\beta}\rangle.
   \end{equation}
   The total radiation is obtained as the latter quantity multiplied by $\hbar \omega$ integrated over frequency. Using Eqs.~(\ref{Eq: Occ num out}) and (\ref{Eq: Unitarity moving1}), one obtains the radiated energy per unit time as
  \begin{align}
   \!\!\! {\cal P}=\int_{0}^{\infty}\!\!\frac{\dd\omega}{2\pi}\, \hbar \omega\!\sum_{\alpha,\beta} (n(\tilde\omega_\alpha,T) -n(\omega,T_{\rm env}) ) \left(\delta_{\beta\alpha }-\left|\bS_{\beta\alpha}(\omega)\right|^2\right).
  \end{align}
  In the absence of motion ($\tilde \omega_\alpha=\omega$), this equation correctly reproduces the thermal radiation from an object out of equilibrium from the environment \cite{Beenakker98,Kruger11}. Interestingly, the moving object radiates energy even when the temperature is zero both in the object and the environment. In this limit, the energy radiation takes the form
  \begin{equation}\label{Eq: Energy Rad lossy T=0}
    {\cal P}=\int_{0}^{\infty}\frac{\dd\omega}{2\pi}\, \hbar \omega\sum_{\alpha,\beta} \Theta(-\tilde\omega_\alpha)\, \left(\left|\bS_{\beta\alpha}(\omega)\right|^2-\delta_{\beta\alpha}\right),
  \end{equation}
  where $\Theta$ is the Heaviside step function. Therefore, spontaneous emission takes place for a process whose frequency $\omega$ is positive in the lab frame while, from the point of view of the moving observer, the corresponding frequency $\tilde\omega_\alpha$ is negative. This mixing between negative and positive frequencies is at the heart of the dynamical Casimir effect \cite{Dalvit11}.

  In Sec. \ref{Sec. Lossy objects: stationary motion}, we employ Eq.~(\ref{Eq: Energy Rad lossy T=0}) to find the spontaneous emission due to a rotating object. Furthermore, we study the configuration of multiple objects in relative motion where we generalize the results presented in this section. In the process, we find that Eqs.~(\ref{Eq: Unitarity moving2}) and (\ref{Eq: Unitarity}) need to be modified for evanescent waves.

  In summary, Eqs.~(\ref{Eq: Enrgy Rad}) and (\ref{Eq: Energy Rad lossy T=0}) express the energy radiation for an accelerating lossless body and a lossy moving object, respectively. In both cases, the radiated energy density is related to the off-diagonal part of the  scattering matrix in $|S|^2$. Hence, they have a characteristic ``Fermi-Golden-Rule'' structure. In fact, to the lowest order, one can think of the off-diagonal $S$-matrix as a \emph{potential} due to the boundary condition or the object's material, akin to the Fermi Golden Rule.

  In the following sections, we provide a variety of examples where we discuss applications of the general formulas presented above.

  \section{Lossless accelerating objects}\label{Sec: Lossless accelerating objects}
  In this section, we consider lossless objects with different shapes in various dimensions undergoing rotational or translational motion or oscillation. In the process, we reproduce some existing results in the literature, and also present many novel applications. Equation (\ref{Eq: Enrgy Rad}) is the central formula according to which we compute and discuss these results.

  \subsection{A Dirichlet point in 1+1d } \label{Sec: point in 1+1}
  The prototype of dynamical Casimir phenomena is the motion of a point-like \emph{mirror} in one dimension \cite{Fulling76,Ford82}. For simplicity, we assume that the ambient vacuum consists of a scalar field, $\Phi$, subject to Dirichlet boundary conditions on the mirror
  \begin{equation}\label{Eq: bc condition 1+1}
    \Phi(t,q(t))=0,
  \end{equation}
  where $q(t)$ is the trajectory of the mirror in time. We use a perturbative scheme \cite{Ford82} where we expand Eq.~(\ref{Eq: bc condition 1+1}) for small $q(t)$ to obtain
  \begin{equation}
    \Phi(t,0)+q(t)\partial_z\Phi(t,0)+\cdots=0\,.
  \end{equation}
  The scattering solution can be formally expanded in powers of $q$,
  \begin{equation}
    \Phi=\Phi_0+\Phi_1+\cdots\,.
  \end{equation}
  The boundary condition, to the first order, takes the form
  \begin{equation}\label{Eq: phi 1}
    \Phi_1(t,0)=-q(t)\partial_z\Phi_0(t,0).
  \end{equation}
  The incoming and outgoing modes are defined as
  \begin{equation}
    \Phi^{\inn/\out}_{\omega}=\sqrt{\frac{c}{{|\omega|}}}\exp[-i \omega (t\pm  z/c)].
  \end{equation}
  Note that the normalization is chosen in accordance with Eqs.~(\ref{Eq: normalization1}) and (\ref{Eq: normalization2}). In the zeroth order, i.e. for a static mirror, we have
  \begin{equation}
    \Phi_0=\Phi^{\inn}_{\omega}-\Phi^{\out}_{\omega}\,.
  \end{equation}
  We can compute $\Phi_1$ by solving the free field equation ($\Box \, \Phi=0$) in the vacuum subject to its time-dependent value at the origin ($z=0$) as given by Eq.~(\ref{Eq: phi 1}). We leave the details to Appendix \ref{App: Plate}; the scattering matrix (from either side of the point) is obtained as
  \begin{equation}
    S_{\omega+\Omega, \omega}=- \frac{ 2i \,\tilde q(\Omega)}{c} \sqrt{{|(\omega+\Omega)}\,{\omega|}}\,,
  \end{equation}
  where $\tilde q(\Omega)$ is the Fourier transform of $q(t)$. (Note that, here and in the following, we only write the off-diagonal correction to the scattering matrix.) One can then compute the radiation according to Eq.~(\ref{Eq: Enrgy Rad}). For a Fourier mode $\Omega$, the integral in the latter equation contributes in the window of $0<-\omega<\Omega$.
  Putting all the pieces together, and multiplying by a factor of two accounting for the scattering from both sides, we find
  \begin{align}\label{Eq: E rad in 1+1}
    P&=\frac{8\hbar}{c^2} \int_{0}^{\infty} \frac{\dd \Omega}{2\pi}|\tilde q(\Omega)|^2
      \int_{-\Omega}^{0}\frac{\dd \omega}{2\pi} \, ({\omega+\Omega})^2 \, |\omega| \nonumber\\
     &=\frac{\hbar}{3\pi c^2} \int_{0}^{\infty} \frac{\dd \Omega}{2\pi}|\tilde q(\Omega)|^2
      \Omega^4\nonumber\\
     &=\frac{\hbar}{6\pi c^2} \int \dd t \, \ddot{q}^2.
  \end{align}
  In the last line, the radiation is expressed as an integral over time.
  From Eq.~(\ref{Eq: E rad in 1+1}), one can infer the dissipative component of the force
  \begin{equation}
    f(t)= \frac{\hbar}{6\pi c^2} \dddot q,
  \end{equation}
  in complete agreement with Ref.~\cite{Ford82}.

  {
    \subsection{Modulated reflectivity in 1+1d}
  For moving bodies, the dynamical Casimir radiation is difficult to detect experimentally since it requires the objects to move at very high frequencies. An alternative approach is suggested by modulating optical properties of a resonant cavity \cite{Braggio05, Agnesi11, Paraoanu11}. In fact, any \emph{linear} time-dependent process can lead to similar dynamical Casimir effects. Modulated reflectivity, for example, generates photons and gives rise to radiation \cite{Dodonov05,Dodonov10}. The latter can be studied within the same framework  that we developed for non-lossy objects\footnote{
  In treating the dynamical Casimir effect in the absence of loss,
  we did not assume that the objects are actually moving.}.  In this section, we consider a point particle in one spatial dimension, but, unlike the model in the previous (sub)section, a linear coupling of (time-dependent) strength $\epsilon$ is introduced at the position of the particle. The field equation then reads
  \begin{equation}
    \left(\frac{1}{c^2}\partial_t ^2-\partial_z^2\right)\Phi(t,z)+\epsilon(t) \delta(z) \Phi(t,0)=0.
  \end{equation}
  We recover a perfectly reflecting object for $\epsilon \to \infty$. An imperfect mirror undergoing arbitrary motion is studied in Ref.~\cite{Fosco07}. Note that in our model the particle is at rest at the origin while the coupling is modulated. The $S$-matrix can be computed by techniques similar to quantum mechanical scattering in a one-dimensional delta potential. For simplicity, we take $\epsilon(t)=\epsilon_0+\epsilon_\Omega \cos(\Omega t)$ with $\epsilon_0\gg\epsilon_\Omega$. We note that there are new scattering channels with incoming waves from one side transmitted to the other side of the object. A scattering ansatz with incoming waves from the RHS is given by
  \begin{align}
    \Phi\approx\begin{cases}
      \Phi_\omega^{R\,\inn}+ r \Phi_\omega^{R\, \out}+ r_{\pm} \Phi^{R\, \out}_{\omega\pm\Omega}, & z>0, \\
       t \Phi_\omega^{L\,\out}+ t_{\pm} \Phi^{L\,\out}_{\omega\pm\Omega}, & z<0,
    \end{cases}
  \end{align}
  where summation is made over both signs, and the wavefunctions denoted by $R(L)$ are defined on the right (left) side. Obviously, the two sets of definitions are related by reversing the sign of the coordinate $z$. In the above ansatz, we have exploited the smallness of the oscillatory part of $\epsilon$ by truncating the sum at the lowest harmonics. One can obtain the scattering amplitudes by matching the functions on the two sides of the mirror while setting the difference in their first derivative to $\epsilon \Phi(t,0)$. We find
  \begin{align}
    r_{\pm}=t_{\pm}=\frac{ i \epsilon_\Omega \sqrt{|\omega (\omega\pm\Omega)|}/c }{(\epsilon_0-2i\omega/c)(\epsilon_0-2i (\omega\pm\Omega)/c)}\,.
  \end{align}
  This equation can be further simplified by assuming $\epsilon_0 \gg \Omega/c$. In this limit, the energy radiation (per unit time) is, according to Eq.~(\ref{Eq: Enrgy Rad}),
  \begin{align}
    {\cal P}= \frac{\hbar \Omega^4\epsilon_\Omega^2}{6\pi c^2\epsilon_0^4}.
  \end{align}
  For a general $\epsilon(t)$ slowly varying around a mean value of $\epsilon_0$, the total energy radiation is given by
  \begin{equation}
    P=\frac{\hbar }{3\pi c^2\epsilon_0^4}\int \dd t \, {\ddot \epsilon(t)}^2.
  \end{equation}
  }

  \subsection{A Dirichlet line in 2+1d}
  Now consider a line extended along the $x$ axis in two spatial dimensions. In this geometry, the incoming and outgoing waves are described by
    \begin{equation}
    \Phi^{\inn/\out}_{\omega \,k_x}=\frac{1}{{\sqrt{k_\perp}}}\exp(-i \omega t+i k_x x\mp  i k_\perp z)\,,
    \end{equation}
  where $k_x$ is the wavevector along the line, and $k_\perp$ is the  perpendicular component, $k_\perp(\omega, k_x)=\sqrt{\omega^2/c^2-k_x^2}$. We assume that the line undergoes a rigid but time-dependent motion, $q(t)$, normal to the $x$ axis. One then finds
  \begin{equation}\label{Eq: S matrix 2+1 line}
    S_{\omega+\Omega k_x, \, \omega k_x}=-2i \, {\tilde q(\Omega)} \sqrt{k_\perp(\omega, k_x) \, k_\perp(\omega+\Omega, k_x)}\,;
  \end{equation}
  see Appendix \ref{App: Plate}. The scattering matrix is diagonal in $k_x$ due to translational symmetry along the $x$-axis. Note that Eq.~(\ref{Eq: S matrix 2+1 line}) is computed only for propagating modes--- evanescent waves fall off rapidly with the distance from the surface and do not contribute to radiation at infinity. The sum over all partial waves in Eq.~(\ref{Eq: Enrgy Rad}) becomes $\int\frac{L {\scriptsize\dd} k_x}{2\pi}$ with $L$ being the extent of the line; the integration is over propagating waves only, i.e. $c|k_x|<|\omega|$ and $c|k_x|< \omega+\Omega$.  Finally, an integration over the frequency, $\omega$, gives
  \begin{align} \label{Eq: E rad 2+1 line}
    P &=\frac{\hbar L}{128 c^3} \int_{0}^{\infty} \frac{\dd \Omega}{2\pi} \, |\tilde q(\Omega)|^2 \Omega^5.
  \end{align}
  In order to write this equation in the time domain, we extend the integral to $(-\infty, \infty )$ and recast the integrand as $\Omega \, |\tilde q(\Omega)|^2 \im\chi(\Omega)$.
  The function $\chi$ is the (normalized) response function whose imaginary part is proportional to the energy dissipation to the environment, $
    \im\chi(\Omega)=\frac{1}{2} \Omega^4 \sgn(\Omega)$.
  Because of causality, the full response function can be obtained via Kramers-Kronig relations. In the time domain, we find this function as
  \begin{align}
  \chi(t)= \frac{24}{\pi}\Theta(t)\pr \frac{1}{t^5}\,,
  \end{align}
  with $\pr$ being the principal part. In this context, the response function relates the force to the object's displacement. So the force acting on the object at time $t$ is given by
  \begin{equation}\label{Eq: force 2+1 line}
    f(t)=\frac{3\hbar L}{16\pi c^3} \int_{-\infty}^{t} \dd t' \pr \frac{1}{(t-t')^5} \, q(t')\,.
  \end{equation}
  The force is manifestly causal, i.e. it depends on $q$ at earlier times. However, Eq.~(\ref{Eq: force 2+1 line}) is possibly divergent near the upper bound of the integral unless a short-time cutoff is introduced to replace this bound by $t-\tau$. This does not affect the dissipative component of the force but regularizes the inertial force which sensitively depends on the large-frequency physics. In fact, in deriving Eq.~(\ref{Eq: force 2+1 line}), we have used the small-frequency behavior of the response function, so this equation should be valid only for large times. Interestingly, the force does not vanish even when the object no longer moves. We find that the force, long after the object comes to a full stop, falls as
  \begin{equation}
    f(t)= \frac{3\hbar L}{16 c^3 t^5} \int \dd t' \,q(t')=\frac{3\hbar L}{16 c^3 t^5 } \,\tilde q(0)\,,
  \end{equation}
  where the integral is over the displacement of the object for the duration of the motion (which is assumed to be much smaller than $t$), and $\tilde q(0)$ is the integrated displacement.

  \subsection{A Dirichlet segment in a waveguide in 2+1d}
\begin{figure}[h]
    \includegraphics[width=60mm]{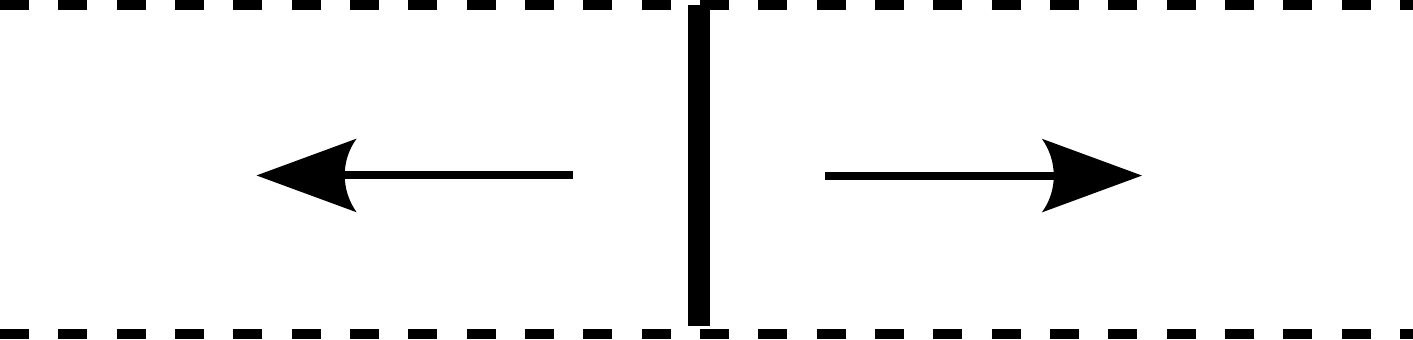}
    \caption{A segment in a waveguide. The arrows indicate the direction along which the segment oscillates. Below a certain frequency, $\omega_{\rm min}=2\omega_0$, the motion is frictionless.}
    \label{Fig: Waveguide}
  \end{figure}

  Next we consider a finite segment of size $L$ along the $x$-axis confined between two infinite Dirichlet lines (a \emph{waveguide}) in two spatial dimensions; see Fig.~\ref{Fig: Waveguide}.
  Dirichlet boundary conditions are assumed on all surfaces. The segment undergoes a rigid motion $q(t)$ parallel to the infinite lines. The only difference compared to the previous (sub)section is that the modes along the $x$-axis are quantized. Therefore the integral over $k_x$ is replaced by a sum over $n$ where $k_n=\frac{n\pi}{L}$. The radiation takes the form (cf. Eq.~(\ref{Eq: E rad 2+1 line}))
  \begin{align}
    P &=\frac{ \hbar }{128c^2} \int_{0}^{\infty} \frac{\dd \Omega}{2\pi}|\tilde q(\Omega)|^2 \Omega^4 \, g\left({\Omega L}/{c}\right),
  \end{align}
  where the function $g$ is defined as
  \begin{align}
    &g(\nu)= \frac{512}{\pi} \sum_{n=1}^{\lfloor \nu/2\pi\rfloor} \nonumber \\ &\int_{\frac{\pi n}{\nu}}^{1-\frac{\pi n}{\nu}} \!\!\!\!\!\! \dd z \, (1-z) \sqrt{(z^2- \frac{\pi^2 n^2}{\nu^2})((1-z)^2- \frac{\pi^2 n^2}{\nu^2})}\,.
  \end{align}
  For large $L$, $g(\nu) \to \nu$, and we recover the results of the previous (sub)section. However, this function vanishes below $\nu=2\pi$; a low-frequency motion does not dissipate energy since propagating waves inside a waveguide have no support in the range $(-\omega_0,\omega_0)$ with $\omega_0$ being the lowest eigenmode of the waveguide, hence $\omega_{\rm min}=2\omega_0= 2\pi c/L$. Close to this frequency, the function $g$ vanishes quadratically,
  \begin{align}\label{Eq: g near 2pi}
    g(\nu)\sim  \frac{8}{\pi^2} \, (\nu-2\pi)^2, \qquad \nu \gtrsim 2\pi .
  \end{align}
  Similar to the previous (sub)section, one can use Kramers-Kronig relations to obtain the response function. Specifically, we are interested in the long-time limit after the object comes to a full stop. The dependence on large $t$ can be inferred from the short-frequency response (Eq.~(\ref{Eq: g near 2pi})) as
  \begin{equation}
    f(t)= -\frac{\hbar }{2\pi c L \,t^3} \re (e^{-i 2 \omega_0 t } \, \tilde q(2\omega_0)),
  \end{equation}
  with $\omega_0=\pi c/L$ as defined before. Note that the force now falls as $1/t^3$ while its amplitude undergoes periodic oscillations at frequency $2\omega_0$, twice the lowest natural frequency of the waveguide.

  \subsection{A Dirichlet plate in 3+1d}
  In this section, we consider a (two-dimensional) plate in three dimensions subject to the same (Dirichlet) boundary conditions.
  Again we assume that the plate undergoes a rigid motion $q(t)$ normal to its surface, and find the scattering matrix as
  \begin{equation}  \label{Eq: S-Mat Plate}
    S_{\omega+\Omega \bk_\|, \, \omega \bk_\|}=-2i \, {\tilde q(\Omega)} \sqrt{k_\perp(\omega, \bk_\|) \, k_\perp(\omega+\Omega, \bk_\|)}\,;
  \end{equation}
  see Appendix \ref{App: Plate}.
  In order to compute the radiation, one should integrate over propagating modes only (both for incoming and outgoing waves), i.e. $c|\bk_\||<|\omega|$ and $c|\bk_\||< \omega+\Omega$. Equation~(\ref{Eq: Enrgy Rad}) then gives
  \begin{align}\label{Eq: E rad 3+1}
    P&=\frac{\hbar L^2}{180\pi^2 c^4} \int_{0}^{\infty} \frac{\dd \Omega}{2\pi}|\tilde q(\Omega)|^2
      \Omega^6\nonumber\\
     &=\frac{\hbar L^2}{360\pi^2 c^4} \int \dd t \dddot{q}^2,
  \end{align}
  where $L^2$ is the area of the plate.
  The resulting (dissipative component of the) force is then
  \begin{equation}
    f(t)=-\frac{\hbar L^2}{360\pi^2 c^4} \, q^{(5)},
  \end{equation}
  again in agreement with Ref.~\cite{Ford82}.

  We also note the difference between odd and even dimensions. In 2 dimensions, the force displays long-time tails, while in 1 and 3 dimensions it is an (almost) instantaneous function of the displacement.

  \subsection{A Dirichlet corrugated plate in 3+1d}
  We can generalize the results in the previous (sub)section by considering a corrugated plate. For corrugations of wavevector $\bq$, the scattering matrix is given by (cf. Eq.~(\ref{Eq: S-Mat Plate}))
  \begin{align}  \label{Eq: S-Mat corrugated Plate}
    S_{\omega+\Omega \bk_\|+\bq, \, \omega \bk_\|}&= \nonumber \\
    -2i \, {\tilde q(\Omega,\bq)} &\sqrt{k_\perp(\omega, \bk_\|) \, k_\perp(\omega+\Omega, \bk_\|+\bq)}\,.
  \end{align}
  where $q$, the displacement from the $x-y$ plane, is a function of both $\omega$ and $\bq$; see Appendix \ref{App: Plate}.
  The condition for propagating waves is modified as $c|\bk_\||<|\omega|$ and $c|\bk_\|+\bq|<\omega+\Omega$.
  The radiation formula should be modified accordingly to ensure that only the propagating modes are integrated.
  While this integral can be computed explicitly, we use a trick as follows: For $\Omega> c|\bq|$, we Lorentz-transform to a frame in which the scattering matrix is diagonal in the wavevector $\bk_\|$; the velocity of this frame is $\bv =\frac{c^2\bq}{\Omega}$. The scattering matrix, the integral measure, as well as the condition for propagating waves are invariant under such a transformation. But the frequency ($\hbar \omega'$ in Eq.~(\ref{Eq: Enrgy Rad})) picks up a factor of $\gamma(\bv)=1/\sqrt{1-\bv^2/c^2}$, while the lower bound of the integral over $\omega$ changes from $-\Omega$ to $-\gamma(\bv)(\Omega-\bv\cdot\bq)=-\Omega/\gamma(\bv)$ which, through comparison with the first line of Eq.~(\ref{Eq: E rad 3+1}), contributes a factor of $1/\gamma^6$. Hence, the radiated energy density in $\Omega$ and $\bq$ becomes
  \begin{equation}
    P(\Omega,\bq)= \frac{\hbar}{360\pi^2 c^4} |\tilde q(\Omega, \bq)|^2\Omega (\Omega^2-c^2\bq^2)^{5/2},
  \end{equation}
  consistent with Ref.~\cite{Golestanian97}. Note that the difference of a factor of two in comparison with Eq.~(\ref{Eq: E rad 3+1}) is in harmony with the setup in Ref.~\cite{Golestanian97} where the plate occupies a half-space. Similar results can be obtained for Neumann boundary conditions \cite{Sarabadani06}.

  \subsection{A Dirichlet sphere in 3+1d}
  In this section,  we consider a sphere subject to Dirichlet boundary conditions linearly oscillating in three dimensions. The oscillation amplitude, $q(t)$, is small compared to the radius of the sphere, $R$. We choose the $z$ axis parallel to the motion and passing through the center of the sphere. The incoming and outgoing waves for a spherical geometry are defined as
  \begin{align}\label{Eq: spherical fns}
    &\Phi^{\inn/\out}_{\omega lm}=\sqrt{\frac{|\omega|}{c}} e^{-i\omega t}\, h_l^{(1,2)}\left(\frac{\omega r}{c}\right)Y_{lm}(\theta, \phi),
  \end{align}
  where $h^{(1,2)}_l$ are spherical Hankel functions, and $Y_{l m}$ is the usual spherical harmonic function.
  Due to azimuthal symmetry, the scattering matrix is diagonal in the index $m$ but possibly mixes different $l$s. The scattering from an oscillating sphere can be computed by using Green's theorem; see Appendix \ref{App: Sphere} for more details. We find the (off-diagonal) scattering matrix as
  \begin{align}
    &S_{\omega+\Omega l'm, \, \omega l m}= \frac{2 i  \tilde q(\Omega)}{c} \, d_{ l l'm}\nonumber \\
    &\times \sqrt{|(\omega+\Omega)\,\omega|} \, F_l\left(\frac{\omega R}{c}\right) F_{l'}\left(\frac{(\omega+\Omega ) R}{c}\right),
  \end{align}
  where $d_{ll'm}$ as defined in Appendix \ref{App: Sphere} is nonzero only for $l'=l\pm1$, and the function $F$ is defined as
  \begin{equation}
    F_l(x)=\frac{1}{x h_l^{(1)}(x)}\,.
  \end{equation}
  Using Eq.~(\ref{Eq: Enrgy Rad}), the radiated energy density is given by
  \begin{align}\label{Eq: E rad sphere sum}
    P(&\Omega)=\frac{8\hbar |\tilde q(\Omega)|^2}{3c^2}\int_{-\Omega}^{0} \frac{\dd \omega}{2\pi} (\omega+\Omega)^2 \omega \nonumber \\
    &\times \sum_{l=0}^{\infty} (l+1)\left|F_l\left(\frac{\omega R}{c}\right)F_{l+1}\left(\frac{(\omega+\Omega ) R}{c}\right)\right|^2.
  \end{align}

  We consider two different limits:

  a) $\Omega R/c \ll 1$. For a slowly oscillating object, we need only consider the lowest partial wave, i.e. the $l=0$ term in Eq.~(\ref{Eq: E rad sphere sum}). One then finds the radiated energy density in frequency as
  \begin{align}
    P(\Omega) = \frac{\hbar \,\Omega^6 R^2}{30\pi c^4}   |\tilde q(\Omega)|^2.
  \end{align}

  b) $\Omega R/c  \gg 1$. In this case, one should include all partial waves up to $l_{\rm max}\approx \Omega R/c\gg1$. Below we closely follow the line of argument in Ref.~\cite{MaiaNeto93}. For large $l$, we have
  \begin{equation}
    |F_l(x)|\approx(1-l^2 /x^2)^{1/4}, \qquad 1\ll l < x.
  \end{equation}
  The sum over all partial waves can then be recast as an integral over $l$ yielding (through the change of variables $\sigma=l/(\Omega R/c)$ and $x=|\omega|/\Omega$)
  \begin{widetext}
  \begin{align}
    P(\Omega)&=\frac{4 \hbar R^2 \Omega^6 |\tilde q(\Omega)|^2 }{3\pi c^4 }  \int_{0}^{1/2} \!\!\!\!\dd\sigma \sigma \int_{\sigma}^{1-\sigma}\!\!\!\!\dd x \,(1-x)x^2 \left[1-\frac{\sigma^2}{x^2}\right]^{1/2}\left[1-\frac{\sigma^2}{(1-x)^2}\right]^{1/2}  =\,\,\,\frac{\hbar \Omega^6 R^2}{270 \pi c^4} \, |\tilde q(\Omega)|^2\,.
  \end{align}
  \end{widetext}
  This equation reproduces the contribution of the TE modes to the electromagnetic version of a perfectly reflecting sphere; see Eq.~(4.20) in Ref.~\cite{MaiaNeto93}.\footnote{There are however mixed terms between the two polarizations which vanish in this limit, see Ref.~\cite{MaiaNeto93}.} Indeed one finds a similar correspondence for a perfectly reflecting plate, that is the radiation due to the TE modes is equal to that of the Dirichlet plate \cite{Neto96}.

  Finally we note that the radiation due to the oscillatory motion  can be computed for a variety of other geometries such as cylinders, ellipsoids, etc.

  \subsection{A spinning object in 2+1d}\label{Sec: A spinning object in 2+1d}
\begin{figure}[h]
   \includegraphics[width=30mm]{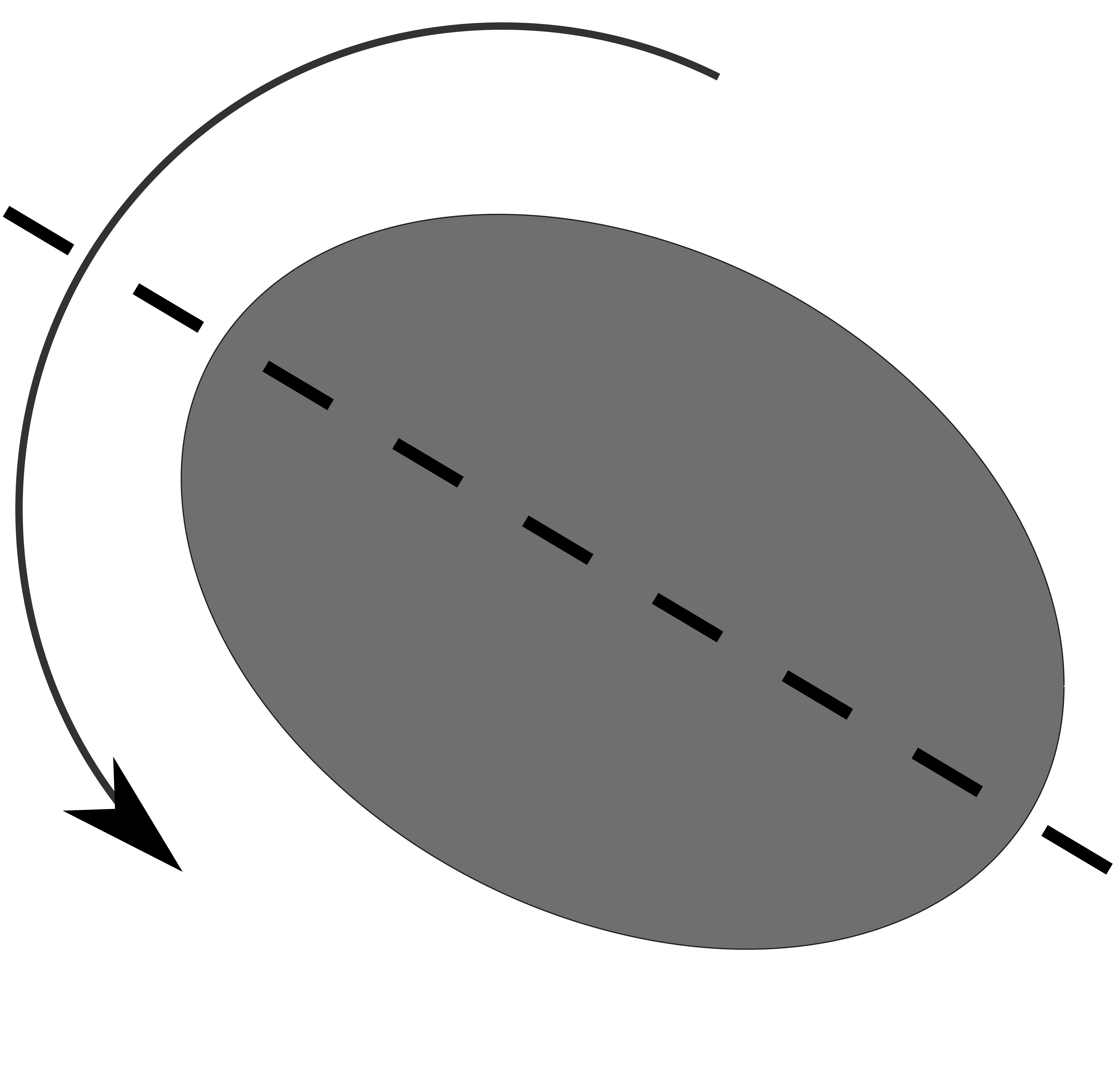}
   \caption{An (asymmetrical) spinning object. The object slows down as it emits ``photons''.} \label{Fig: spinning}
  \end{figure}

  Heretofore, we studied examples where the object is accelerated by an external force. In this section, we consider an object rotating at a constant angular velocity $\Omega$. We further assume that the object is not rotationally symmetric, as illustrated in Fig. \ref{Fig: spinning}.
  As usual, we assume a scalar field subject to Dirichlet boundary conditions on the object's surface, and limit ourselves to 2+1 dimensions.

  Since the orientation changes with time, waves impinging on the object are partially scattered at a shifted frequency determined by $\Omega$. The scattering matrix is more conveniently computed by going to the object's reference frame.

  We start with the field equations in the laboratory (static) frame. The wave equation, $\Box\Phi=0$, in polar coordinates is
   \begin{equation}
     \left[\frac{1}{c^2}\partial_t ^2 -\frac{1}{r}\partial_r r\partial_r -\frac{1}{r^2}\partial_\phi^2\right]\Phi(t,r, \phi)=0.
  \end{equation}
  The incoming and outgoing waves are defined as
  \begin{equation}\label{Eq: wavefunction 2+1}
    \Phi^{\inn/\out}_{\omega m}(t,r, \phi)= e^{-i \omega t} e^{i m \phi} H_m^{(1,2)}\left(\frac{\omega r}{c}\right)\,,
  \end{equation}
  where $H$ is the Hankel function. (We have dropped an irrelevant constant in the definition of the these functions.)
  The rotating frame is described by
  \begin{equation}\label{Eq: Coordinate transf}
    t'=t, \quad r'=r, \quad \phi'=\phi-\Omega t.
  \end{equation}
  The field equation in the latter frame takes the form
  \begin{equation}\label{Eq: e.o.m. rotating frame}
    \left[\frac{1}{c^2}(\partial_t -\Omega \partial_{\phi'})^2 -\frac{1}{r}\partial_r r\partial_r -\frac{1}{r^2}\partial_{\phi'}^2\right]\Phi'(t,r,\phi')=0.
  \end{equation}
  Note that $\Phi'(t,r,\phi')=\Phi(t,r, \phi)$. Specifically, in the new coordinate system, the functions $\Phi_{\omega m}$ as defined in Eq.~(\ref{Eq: wavefunction 2+1}) become
  \begin{equation}
    \Phi'_{\omega-\Omega m \, m}(t,r, \phi')= e^{-i(\omega-\Omega m)t} e^{i m \phi'} H_m^{(1,2)}\left(\frac{\omega r}{c}\right).
  \end{equation}
  The rotating frame is more convenient to write the scattering ansatz as the object does not move in this frame, and thus the time dependence drops out as a phase factor. In the latter frame, the boundary conditions take the form
  \begin{align}
    &\Big[e^{im \phi'} H_m^{(2)}\left(\frac{\omega r}{c}\right)\nonumber \\ \!\!\!\!\!+\sum_{m'} & S_{m',m}e^{i m'\phi'}H^{(1)}_{m'}\left(\frac{(\omega-\Omega(m-m'))r}{c}\right)\Big]_{\Sigma}=0,
  \end{align}
  where $\Sigma$ denotes the boundary. One can see that this equation indeed satisfies Eq.~(\ref{Eq: e.o.m. rotating frame}) once the time dependence, $e^{-i(\omega-\Omega m)t}$, is restored.
  The scattering matrix sends the frequency $\omega$ to $\omega-\Omega(m-m')$ from the point of view of an observer in the lab frame.
  To obtain an analytical expression for the scattering matrix, we consider the non-relativistic limit where the object's (linear) velocity is small compared to $c$. It then suffices to compute the scattering matrix for the lowest partial waves. As a specific example we consider an ellipse close to a circle of radius $R$ with $\delta$ being the difference of the two semiaxes, i.e., in polar coordinates, defined as $r(\phi)=R+\cos(2\phi)\delta/2$. The scattering matrix to the lowest order in $\delta$ is then obtained as
  \begin{align}\label{Eq: S mat revolving ellipse}
    S_{ \omega+2\Omega\, 2 , \, \omega\,0}&=\frac{i\pi (\omega+2\Omega)^2 R \delta}{8c^2\log(|\omega| R/c)}\,\,, \nonumber \\
    \quad S_{ \omega+2\Omega \, 0,\,  \omega \, -2}&=\frac{i\pi\omega^2 R \delta}{8c^2\log((\omega+2 \Omega )R/c)}\,\,.
  \end{align}
  The energy radiation per unit time, according to Eq.~(\ref{Eq: Enrgy Rad}), is
  \begin{align}\label{Eq: En Rad spinning ellipse}
    {\cal P}    &\approx\frac{\pi\hbar R^2\delta^2\Omega^6 }{10 c^4\log(\Omega R/c)^2}\,.
  \end{align}
  Therefore, an (asymmetrical) object which is spinning, even at a constant rate, slows down due to quantum dissipation. Note that the torque (due to the back-reaction) is simply the radiation rate divided by the frequency $\Omega$.

  \subsection{A Dirichlet disk in 2+1 dimensions: linear vs angular motion}
\begin{figure}[h]
   \includegraphics[width=45mm]{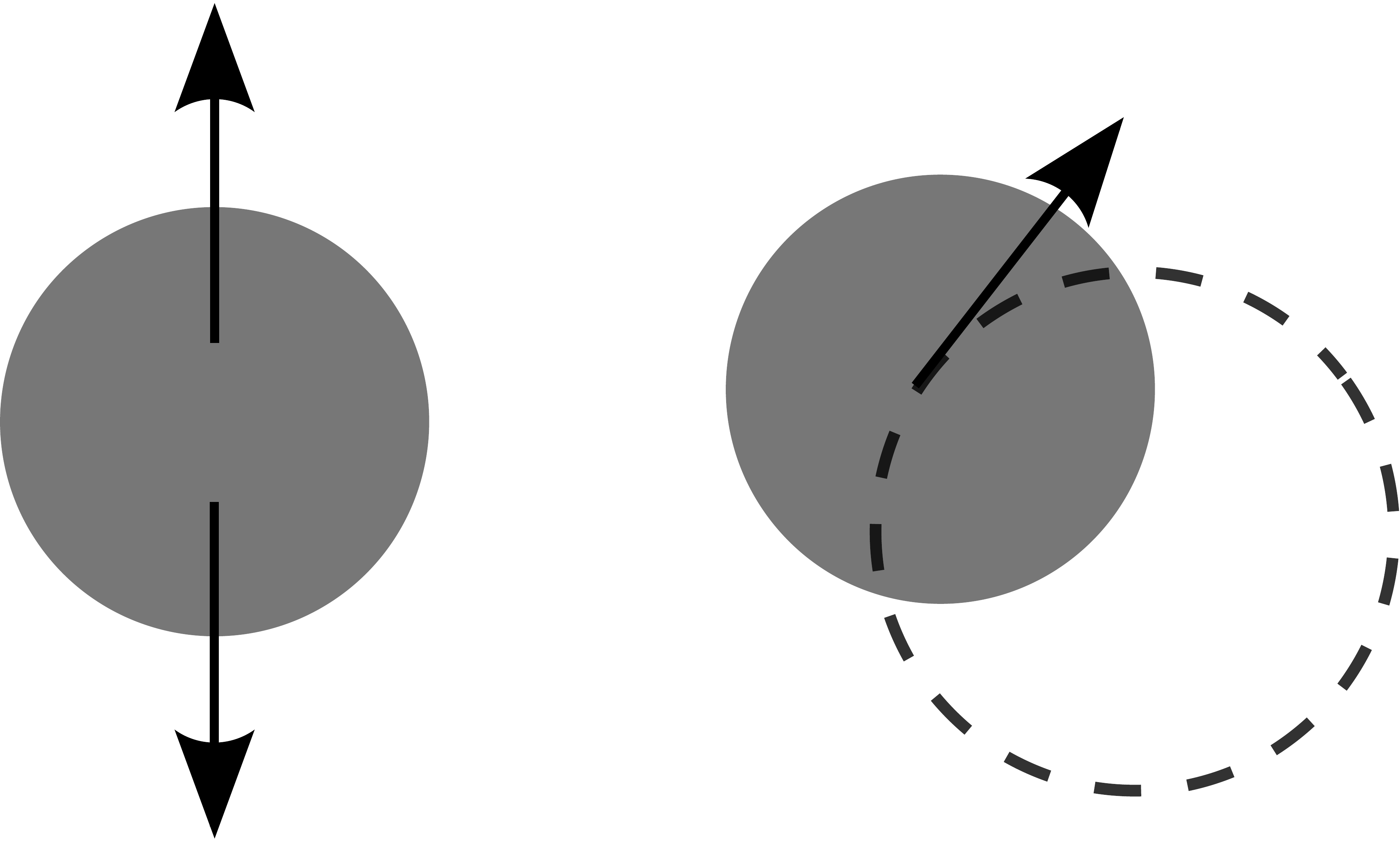}
   \caption{Linear {\it vs} angular motion; the radiated energy is comparable in the two cases.} \label{Fig: 2D}
  \end{figure}

  Now consider a circular disk of radius $R$ subject to Dirichlet boundary conditions in two spatial dimensions. Below we contrast two different types of motion, see Fig.~\ref{Fig: 2D}.
  First we consider a linear oscillation, $q(t)=\delta \cos\Omega t$ (with $\delta$ being the amplitude of the oscillation) along the $x$-axis.
  The scattering from an oscillating disk can be obtained by using Green's theorem; see Appendix \ref{App: Disk} for more details. We obtain the scattering matrix as
  \begin{equation}
    S_{\omega+\Omega \, m\pm 1, \omega \, m}=\frac{2 i \, \delta }{\pi R} \, M_{m}\left(\frac{\omega R}{c}\right) M_{m\pm1}\left(\frac{(\omega+\Omega) R}{c}\right)\!,
  \end{equation}
  where the function $M$ is
  \begin{equation}
    M_m(x)=\frac{1}{H_m^{(1)}(x)}\,.
  \end{equation}
  At low frequency, $\Omega R/c \ll 1$, the scattering matrix for the lowest partial waves is obtained as
  \begin{align}\label{Eq: S mat oscillating disk}
    S_{ \omega+\Omega\, \pm 1, \omega\, 0}&=\mp \frac{i\pi (\omega+\Omega)\delta}{4 c\log(\omega R/c)}\,.
  \end{align}
  The energy radiation (per unit time) can be computed from Eq.~(\ref{Eq: Enrgy Rad}); for a linearly oscillating disk, we find
  \begin{equation}\label{Eq: P-lin}
    {\cal P}_{\updownarrow}= \frac{\pi \hbar \,\delta^2 \Omega^4 }{64c^2 \log(\Omega R/c)^2}\,.
  \end{equation}

  Next we consider the same disk undergoing orbital motion. The latter can be thought as a rotation around a point off the center of the disk. Specifically, we assume that the disk's center undergoes a trajectory $(r, \phi)=(\delta, \Omega t)$ in polar coordinates, i.e. the center is at a fixed radius $\delta$ while orbiting around the origin at frequency $\Omega$. The scattering process can be examined by techniques similar to the previous (sub)section where we considered a spinning object. We must assume $\Omega R/c\ll 1$ as well as $\delta\ll R$---similar to the linear motion. The latter allows us to compute the scattering matrix only for the lowest partial waves, as
  \begin{align}\label{Eq: S mat revolving disk}
    S_{ \omega+\Omega\, 1, \omega \, 0}&=-\frac{i\pi (\omega+\Omega)\delta}{2 c\log(|\omega| R/c)}\,, \nonumber \\
    S_{ \omega+\Omega \, 0,  \omega\, -1}&=\frac{i\pi\omega\delta}{2c\log((\omega+\Omega )R/c)}\,.
  \end{align}
  Note the similarity between Eqs.~(\ref{Eq: S mat oscillating disk}) and (\ref{Eq: S mat revolving disk}).
  The radiated energy by a revolving disk can then be computed according to Eq.~(\ref{Eq: Enrgy Rad}), as
  \begin{equation}\label{Eq: P-ang}
    {\cal P}_{\circ}=\frac{\pi\hbar \delta^2 \Omega^4}{24 c^2\log(\Omega R/c)^2}\,.
  \end{equation}
  It is interesting to compare Eqs.~(\ref{Eq: P-lin}) and (\ref{Eq: P-ang}) where linear and angular motion have been considered, respectively. We notice that, apart form a proportionality constant, the analytical form of the dissipated energy is identical in the two cases.

  The above results can be compared with the previous (sub)section. In the former case, the setup is symmetric under inversion with respect to the origin while the latter is not. Thus, in the lowest order, scattering of waves changes the angular momentum by two units for a spinning ellipse and by one unit for a disk in (linear or circular) motion; cf. Eqs.~(\ref{Eq: S mat revolving ellipse}), (\ref{Eq: S mat oscillating disk}) and (\ref{Eq: S mat revolving disk}). Consequently, the energy radiation in the former case, Eq.~(\ref{Eq: En Rad spinning ellipse}), is suppressed by two orders of magnitude in comparison with Eqs.~(\ref{Eq: P-lin}) and (\ref{Eq: P-ang}) for a disk.

  \section{Stationary motion of lossy objects} \label{Sec. Lossy objects: stationary motion}
  In this section, we consider lossy objects but limit ourselves to constant (linear or angular) speed. Among other results, we reproduce existing formulas in the literature with significantly less labor. Our starting point is Eq.~(\ref{Eq: Energy Rad lossy T=0}) in application to rotating objects. We then generalize to the case of multiple objects in relative motion.

  \subsection{Rotating object}
  Let us consider a solid of revolution spinning around its axis of symmetry at frequency $\Omega$. We choose polar coordinates $(r, \phi,z)$ where the $z$ direction is along the axis of symmetry. The latter coordinates describe the lab frame in which the object is rotating. The rotating (or comoving) frame is defined by the coordinate transformation
  \begin{equation}
    t'=t, \quad r'=r, \quad \phi'=\phi-\Omega t, \quad z'=z.
  \end{equation}
  A partial wave defined by frequency $\omega$ and azimuthal index $m$ in the lab frame is characterized by the frequency
  \begin{equation}\label{Eq: omega co-moving}
    \tilde \omega_m=\omega-\Omega m
  \end{equation}
  from the perspective of the rotating frame; see Ref.~\cite{Maghrebi12} for a detailed discussion. In harmony with the discussion in Sec. \ref{Sec: Formalism}, the object spontaneously emits energy when $\omega>0$ and $\tilde \omega_m<0$, i.e. in the frequency window
  \begin{equation}
    0<\omega<\Omega m,
  \end{equation}
  or the so-called superradiating regime first introduced by Zel'dovich \cite{Zel'dovich71}.
  Therefore, the energy radiation per unit time from a rotating object to the environment at zero temperature is given by
  \begin{equation}\label{Eq: lossy Rotating}
    {\cal P}=\int_{0}^{\infty} \frac{\dd\omega }{2\pi}\, \hbar \omega \, \tr\left[\,\Theta(\Omega \, \hat l_z-\omega)\left(\bbS(\omega)\,\bbS(\omega)^\dagger-\bI\right)\right],
  \end{equation}
  where $\hat l_z$ is the $z$-component of the angular momentum operator in units of $\hbar$. Note that the scattering matrix is diagonal in frequency, $\omega$, since the object is undergoing a stationary motion with its shape and orientation fixed in time.
  Equation (\ref{Eq: lossy Rotating}) indeed gives the spontaneous  emission by a rotating object consistent with Ref. \cite{Maghrebi12} where the Rytov formalism is used through an involved analysis. For a small object, only the lowest partial waves contribute to the radiation, and we recover the results of Ref.~\cite{Manjavacas10}. Note that in deriving Eq.~(\ref{Eq: lossy Rotating}) we did not use any approximations regarding the velocity of the rotating object.

  For the sake of generalization to multiple objects, we point out that Eq.~(\ref{Eq: 1}) can be interpreted in a simple way to arrive at the same results. According to this equation, the probability amplitude for spontaneous emission is given by
  \begin{equation}
    {\cal A}_m(\omega)= U_m(\omega),
  \end{equation}
  where we have suppressed all quantum numbers other than $\omega$ and $m$, and used the fact that the amplitude is diagonal in $m$ due to the rotational symmetry of the object. The rate of this process is
  \begin{equation}
    {\cal N}_m(\omega)=|U_m(\omega)|^2=|S_m(\omega)|^2-1,
  \end{equation}
  where, in the last equality, we have used Eq.~(\ref{Eq: Unitarity moving1}) in the superradiating regime, i.e. for $0<\omega<\Omega m$. Note that, in this regime, $|S_m(\omega)|>1$, hence superradiance. The integral of $\cal N$ multiplied by $\hbar \omega$ (over superradiating frequencies) reproduces the energy radiation as given by Eq.~(\ref{Eq: lossy Rotating}).

  \subsection{Moving plates}
  A system comprising two lossy parallel plates undergoing relative lateral motion is the canonical example of non-contact friction. In the following, we sketch a simple derivation of this friction based on Eqs.~(\ref{Eq: 1}) and (\ref{Eq: Unitarity}).\footnote{For a uniform translational motion, one can Lorentz-transform Eq.~(\ref{Eq: Unitarity}) to the moving frame.} We first note that Eq.~(\ref{Eq: Unitarity}) can be interpreted via a classical argument. One can denote the RHS of this equation as the rate of ``photon'' absorption in a dispersive medium. Current conservation implies that the latter should be equal to the influx of the field quanta outside the body. Let us consider a classical wave scattered from the object as
  \begin{equation}
    \Phi=\Phi^{\inn}_\alpha+\sum_\beta S_{\beta\alpha}\Phi^{\out}_\beta,
  \end{equation}
  where we have suppressed the frequency, $\omega$.
  The current density going into the body is given by
  $ \frac{-1}{2i}\left[{\Phi^{*}} \, \nabla {\Phi}-\nabla {\Phi^*} \, {\Phi}\right]$.
  Since $\Phi_\alpha$ and $\Phi_\beta$ are properly normalized (see Eqs.~(\ref{Eq: normalization1}) and (\ref{Eq: normalization2}) and the explanation thereafter), the total influx of field quanta is
  \begin{equation}
    \frac{i}{2}\oint \dS \cdot \left[{\Phi^{*}} \, \nabla {\Phi}-\nabla {\Phi^*} \, {\Phi}\right]=1- \sum_{\beta}|S_{\beta\alpha}|^2.
  \end{equation}
  But this is exactly the LHS of Eq.~(\ref{Eq: Unitarity}).

  To study moving plates we need to extend Eq.~(\ref{Eq: Unitarity}) to evanescent waves which arise in non-compact geometries (plates, cylinders, etc.), but are absent for a compact geometry. Since such waves are not propagating, ``incoming'' and ``outgoing'' wave functions lose their straightforward interpretation. In other words, they do not carry currents
  \begin{align}\label{Eq: evanescent wave normalization 1}
  \oint \dS \cdot \left[{\Phi^{\out/\inn*}_{\alpha}} \, \nabla {\Phi^{\out/\inn}_{\beta}}-\nabla {\Phi^{\out/\inn*}_{\alpha}} \, {\Phi^{\out/\inn}_{\beta}}\right] =0,
  \end{align}
  but satisfy a different relation (after proper normalization)
  \begin{equation}\label{Eq: evanescent wave normalization 2}
  \frac{1}{2}\oint \dS \cdot \left[{\Phi^{\inn*}_{\alpha}} \, \nabla {\Phi^{\out}_{\beta}}-\nabla {\Phi^{\inn*}_{\alpha}} \, {\Phi^{\out}_{\beta}}\right] =\delta_{\alpha \beta}.
  \end{equation}
  Therefore an incoming wave scattered from the object, $\Phi=\Phi^{\inn}_\alpha+\sum_\beta S_{\beta\alpha}\Phi^{\out}_\beta$, carries an influx of ``photons'' given by
  \begin{equation}
    \frac{i}{2}\oint \dS \cdot \left[{\Phi^{*}} \, \nabla {\Phi}-\nabla {\Phi^*} \, {\Phi}\right]= 2 \im S_{\alpha\alpha},
  \end{equation}
  where Eqs.~(\ref{Eq: evanescent wave normalization 1}) and (\ref{Eq: evanescent wave normalization 2}) are used.
  Then the conservation of current dictates
  \begin{equation}\label{Eq: Unitarity evanescent}
    (\bbU \bbU^\dagger )_{\alpha\alpha}=2 \im \left( \bbS\right)_{\alpha\alpha}.
  \end{equation}
  Equations (\ref{Eq: Unitarity}) and (\ref{Eq: Unitarity evanescent}) define the matrix $\bbU$ in terms of the scattering (or reflection) matrix for propagating and evanescent waves, respectively.

  We can now compute the friction between two ``dielectric'' plates moving in parallel. We assume that the first plate is at rest while the other plate, separated from the first by a distance $d$, moves at a constant velocity $v$ along the $x$ axis. Because of translational symmetry, all matrices are diagonal in the frequency $\omega$ and the wavevector $\bk_\|$ parallel to the surface.
  Here, Eq.~(\ref{Eq: 1}) finds a two-fold application. On one hand, it allows for spontaneous emission from an object, while on the other hand, it describes the reflection and absorption of waves by a second object.

  1) Spontaneous emission: the source fluctuations in the first plate give rise to outgoing wave fluctuations. The amplitude of the spontaneous emission is given by
  \begin{equation}
    {\cal A}_1=U_1,
  \end{equation}
  where the dependence of the matrix $\bU$ on $\omega$ and $\bk_\|$ is implicit.
  Note that incoming waves do not contribute to spontaneous emission.

  2) Reflection: These outgoing waves propagate to the second plate and get a factor of $e^{i k_\perp d}$ with $k_\perp=\sqrt{\omega^2/c^2-\bk_\|^2}$; a phase factor for propagating waves while exponentially decaying for evanescent waves. There they are partly reflected and partly absorbed by the second plate.
  The amplitude for ``photons'' spontaneously emitted by the first plate and then absorbed by the second one is
  \begin{equation}
    {\cal A}_{2 \leftarrow 1}=e^{i k_\perp d}U_2U_1.
  \end{equation}
  Equivalently, the rate of the latter process is given by
  \begin{equation}\label{Eq: 2 <-- 1}
  \mathcal N^{1^{\scriptsize\rm st}}_{2 \leftarrow 1}=|{\cal A}_{2 \leftarrow 1}|^2 \, n_1=  \left|e^{i k_\perp d}\right|^2 |\bU_{2}|^2  |\bU_{1}|^2 \, n_1,
  \end{equation}
  where $n_1=n(\omega, T_1)$ is the Bose-Einstein occupation number defined at temperature $T_1$. The superscript $1^{\scriptsize\rm st}$ indicates that Eq.~(\ref{Eq: 2 <-- 1}) is computed within the first reflection. One can similarly compute $\mathcal N_{1 \leftarrow 2}$, the current from the second to the first plate. But in the latter case, $n_2(\omega,\bk_\|)$ is centered at $\omega-v k_x$, i.e. $n_2=n(\omega-v k_x,T_2)$, because  the thermal fluctuations are defined with respect to the comoving frame.\footnote{For relativistic velocities, one should also include the Lorentz factor $\gamma(v)=1/\sqrt{1-v^2/c^2}$.}
  The total flux from the first to the second plate, within the first reflection, is then
  \begin{equation}
   \mathcal N^{1^{\scriptsize\rm st}}_{2 \leftarrow 1}-\mathcal N^{1^{\scriptsize\rm st}}_{1 \leftarrow 2}= \left|e^{i k_\perp d}\right|^2 |\bU_{2}|^2  |\bU_{1}|^2 \, (n_1-n_2).
  \end{equation}
  One can easily sum the contributions from multiple reflections,
  \begin{align}
    {\cal A}_{2 \leftarrow 1}^{\rm tot}
    &=\sum_{n=0}^{\infty} e^{i k_\perp d}U_2 \, \left( e^{2 i k_\perp d} R_1 R_2 \right)^n \,U_1\nonumber \\
    &=\frac{e^{i k_\perp d}U_2U_1}{1-e^{2 i k_\perp d} R_1 R_2 }\,,
  \end{align}
  where $R_1$ and $R_2$ are the reflection matrices. Note that the $n$-th term in the last equation is the amplitude for a ``photon'' spontaneously emitted by the first plate ($U_1$), reflected $n$ times from the two plates ($( e^{2 i k_\perp d} R_1 R_2)^n$) before finally getting absorbed by the second plate ($U_2$). The amplitude $ {\cal A}_{1 \leftarrow 2}^{\rm tot}$ is obtained similarly.
  The total rate then becomes
  \begin{equation}
    \mathcal N_{2 \leftarrow 1}-\mathcal N_{1 \leftarrow 2}= \frac{\left|e^{i k_\perp d}\right|^2 |\bU_{2}|^2  |\bU_{1}|^2}{|1-e^{2ik_\perp d} R_1 R_2|^2} \, (n_1-n_2).
  \end{equation}
  Also note that, from Eqs.~(\ref{Eq: Unitarity}) and (\ref{Eq: Unitarity evanescent}), we have
  \begin{equation}\label{Eq: U vs S}
    |\bU_i|^2=\begin{cases}
      1- |R_i|^2,  & {\mbox{for propagating waves}},\\
      2 \im R_i\,, & {\mbox{for evanescent waves}}.
    \end{cases}
  \end{equation}
  Finally friction is the rate of (lateral-)momentum transfer integrated over all partial waves,
  \begin{equation}\label{Eq: friction}
    f=\int_{0}^{\infty} \frac{\dd\omega}{2\pi}\int \frac{L^2\dd\bk_\|}{(2\pi)^2} \,\,\hbar k_x \,\frac{\left|e^{i k_\perp d}\right|^2 |\bU_{2}|^2  |\bU_{1}|^2}{|1-e^{2ik_\perp d} R_1 R_2|^2} \, (n_1-n_2).
  \end{equation}
  We should emphasize that the reflection matrix for the second plate should be computed in its rest-frame, and then transformed to the lab frame according to Lorentz transformations.
  The last equation is the analog of the results in Refs.~\cite{Pendry97,Volokitin99} for the scalar field.

  To be more concrete, we consider a scalar model that is
  described by a free field equation in empty space while inside the object a ``dielectric'' (or, a response) function $\epsilon$ is assumed which characterizes  the object's dispersive properties. The field equation for this model reads
  \begin{equation}\label{static Lagrangian}
  \left(\epsilon(\omega,\bx) \omega^2 +\nabla^2 \right)\Phi(\omega,\bx)=0,
  \end{equation}
  with $\epsilon$ being 1 in the vacuum, and a frequency-dependent constant inside the object.

  For a semi-infinite plate, the reflection matrix $R$ is given by
  \begin{eqnarray}\label{Eq: reflection matrix}
    R_{\omega \bk_\|}=-\frac{\sqrt{\epsilon \, \omega^2/c^2-\bk_\|^2}-\sqrt{\omega^2/c^2-\bk_\|^2}}{\sqrt{\epsilon \, \omega^2/c^2-\bk_\|^2}+\sqrt{\omega^2/c^2-\bk_\|^2}}.
  \end{eqnarray}
 This is easily obtained by solving the field equations inside and outside the plate and demanding the continuity of the field and its first derivative along the boundary. In a moving frame, the frequency and the wavevector should be properly Lorentz-transformed.

 These reflection matrices can then be inserted in Eq.~(\ref{Eq: friction}) to compute the frictional force.

  \subsection{An \emph{atom }moving parallel to a plate}
  In this section, we consider a small spherical object, an \emph{atom}, moving parallel to a plate. In the non-retarded limit, an electrostatic computation is done in Ref.~\cite{Barton10} for a similar setup. For our purposes, it is more convenient to consider the rest frame of the ``atom'' in which the plate moves laterally. This is another example of stationary motion where the geometrical configuration does not change even though the objects are undergoing relative motion. We assume a small spherical object of radius $a$ (much smaller than the separation distance $d$), such that the first-reflection approximation suffices. Similar to the previous (sub)section, we first consider spontaneous emission by each object. The plate (denoted by sub-index 2) emits ``photons'' of frequency $\omega$ and wavevector $\bk_\|$ with a probability amplitude
  \begin{equation}
    {\cal A}_2(\omega,\bk_\|)=U_2(\omega,\bk_\|).
  \end{equation}
  Then, these waves propagate to and reflect from the ``atom.'' Planar waves pick up a factor $e^{i k_\perp d}$ upon traveling a distance $d$. To find the scattering off of the spherical object, we must change to a basis of spherical waves. A planar wave can be expressed as a superposition of spherical waves as
  \begin{equation}
    e^{i \bk\cdot \bx}=4\pi\sum_{lm} i^l j_l ({\omega r}/{c})  Y_{lm}(\hat \bx)\,    Y_{lm}^*(\hat \bk),
  \end{equation}
  where $\hat \bx$ and $\hat\bk$ are the unit vectors parallel to the vectors $\bx$ and $\bk$, respectively. A planar wave, $\Phi_{\omega \bk_\|}$, defined with respect to a reference point on the plate's surface below the sphere's center is related to spherical waves, $\Phi_{\omega l m}$, centered around the ``atom'' as
  \begin{equation}
    \Phi_{\omega \bk_\|}^{\out}=\sum_{lm}\frac{2 \pi  i^l e^{ik_\perp d} Y_{lm}^*(\hat \bk)}{\sqrt{k_\perp \omega/c}} \left(\Phi_{\omega l m}^{\inn}+\Phi^{\out}_{\omega l m}\right);
  \end{equation}
  see Appendix \ref{App: 1} for the definition of planar and spherical waves. Then the amplitude for ``photons'' spontaneously emitted by the plate and then absorbed by the sphere is
  \begin{equation}
    {\cal A}_{1 \leftarrow 2}=\frac{2 \pi  i^l e^{ik_\perp d} Y^*_{lm }(\hat \bk)}{\sqrt{k_\perp \omega/c}}  \, U_{1\, lm}(\omega) U_2(\omega,\bk_\|),
  \end{equation}
  where $U_1$ characterizes the loss due to the ``atom.'' Similarly, we can compute the amplitude ${\cal A}_{2 \leftarrow 1}$ for the inverse process where the spontaneous emission due to the ``atom'' is absorbed by the plate. One can then obtain the rate of energy or momentum transfer between the objects. An analysis similar to the previous (sub)section gives the force within the first reflection,
  \begin{align}
    &f=\int_{0}^{\infty}\frac{\dd\omega}{2\pi}\int\frac{\dd\bk_\|}{(2\pi)^2} \hbar k_x \, (n_1-n_2)\nonumber \\
    & \times \sum_{l,m} \frac{\left|e^{i k_\perp d}\right|^2 |Y_{lm}(\hat \bk)|^2 |U_{1\, lm}(\omega)|^2 |U_2(\omega,\bk_\|)|^2 }{|k_\perp| \omega/4\pi^2 c} \, ,
  \end{align}
  where $n_1(\omega)=n(\omega,T_1)$ and $n_2(\omega,\bk)=n(\omega-v k_x,T_2)$ are the Bose-Einstein factors for the ``atom'' at temperature $T_1$ and the plate at temperature $T_2$, respectively. Note that we have only considered the first reflection as the ``atom'' is small compared to the separation distance.
  The matrix $U_2$ is given as in Eq.~(\ref{Eq: U vs S}) while, for the spherical object, there is no evanescent wave and thus the matrix $U_1$ is constrained by
  \begin{equation}
    |U_{1\, lm}(\omega)|^2=1-|S_{lm}(\omega)|^2,
  \end{equation}
  with $S_{ lm}(\omega)$ being the scattering matrix of the ``atom.'' This equation indicates that a frictional force (or energy transfer) arises only if the ``atom'' is lossy, i.e. $|S_{lm}(\omega)|<1$.
  For the scalar model introduced in the previous (sub)section, the sphere's scattering matrix is
  \begin{equation}\label{Eq: scattering matrix sphere}
    S_{l m}(\omega)=-\frac{h_l^{(2)}(\omega a/c) \partial_a {j_l}(n\omega a/c)-\partial_a{h_l^{(2)}}(\omega a/c) \, {j_l}(n\omega a/c) }{ h_l^{(1)}(\omega a/c)\partial_a{j_l}(n\omega a/c)-\partial_a{h_l^{(1)}}(\omega a/c)\, {j_l}(n\omega a/c) },
  \end{equation}
  where $a$ is the sphere's radius, and $n(\omega)=\sqrt{\epsilon_S(\omega)}$ with $\epsilon_S$ being the ``dielectric'' function of the spherical object. To the lowest order in $a/d$, we shall limit ourselves to the low-frequency scattering of the partial wave $l=m=0$.\footnote{One should note that in the case of electrodynamics there are no monopole fluctuations and thus the leading contribution to the friction comes from $l=1$ \cite{Barton10}.} Within this approximation, the friction at zero temperature takes the form
  \begin{equation}
    f\approx \frac{4\hbar a^3}{3\pi^2c^2}\int_{k_x>0} \!\!\dd \bk_\| \int_{0}^{v k_x} \!\! \dd \omega \frac{e^{-2|\bk_\||d} k_x \omega^2 \im \epsilon_S(\omega) \im R_{\omega' \bk'_\|}}{|\bk_\||}\,.
  \end{equation}
  The reflection matrix $R$ can be obtained from Eq.~(\ref{Eq: reflection matrix}) via Lorentz transformation.
  Similarly, one can consider the frictional force between a rotating sphere and a stationary plate \cite{zhao12}. With our scalar model, the scattering matrix for a rotating sphere is obtained from Eq.~(\ref{Eq: scattering matrix sphere}) by changing the argument of the Bessel functions to $n(\tilde \omega_m) \tilde \omega_m a/c$ where $\tilde \omega_m=\omega-\Omega m$. Having the scattering matrices of a moving plate and a rotating sphere, one can compute the friction when both objects are set in motion.

  \section{Summary and outlook}
  In this work, we have developed a unified scattering approach to the dynamical Casimir effect. We have obtained general formulas for the radiation from moving objects for accelerating boundaries and modulated optical devices without loss, as well as lossy bodies in uniform motion. We provided and studied numerous examples, many of which are novel, to better illustrate the technical power and conceptual elegance of the scattering approach.

  Extensions to more realistic boundary conditions and field theories should be of both theoretical and practical interest. A superfluid liquid, for example, presents a natural framework to study the effects of motion in a quantum vacuum. Indeed, similar effects such as radiation and spontaneous emission have been discussed for a superfluid \cite{Volovik99}.

  Quantum electrodynamics requires a further treatment due to its vector nature. An obvious complication arises since electromagnetic waves are polarized, enlarging the scattering matrix, and complicating practical computations.
  Realistic materials also include loss. In Sec.~\ref{Sec: Lossless accelerating objects}, we considered accelerating objects with perfect boundary conditions. Extensions to lossy objects require a generalization of the formalism presented in Sec.~\ref{Sec: Formalism}. An expression solely in terms of the scattering matrix is desired in the latter case when object's acceleration and dispersive properties  interplay in a rather complex fashion. One can anticipate that the computation of the scattering matrix will be more complicated in this case. It is also worthwhile to consider configurations of multiple objects in arbitrary motion. We have partially tackled this problem in the context of stationary motion in Sec.~\ref{Sec. Lossy objects: stationary motion}, while extensions to accelerating objects will present new challenges and provide further insights. Specifically, one can ask how the (inertial as well as dissipative) forces between two objects change as the result of their motion or acceleration.

  The formulation of the dynamical Casimir effect in terms of the scattering matrix should also provide an efficient prescription for numerical computations. The scattering matrix is purely a classical quantity, and presumably can be numerically computed with high precision. This is particularly important if the motion cannot be treated perturbatively---when the speed, the amplitude of oscillations, or the corrugations of boundaries are not small. Even in these cases, the scattering formalism is applicable, and numerical methods should prove useful.

  In the light of recent experiments on dynamical Casimir effect, precise computations of the effect of geometry and motion are needed. We believe that our formalism provides an efficient analytical, and possibly even numerical, computational tool.

  \section*{Acknowledgements}
  We have benefitted from discussions with R. L.~Jaffe, G.~Bimonte, T.~Emig, V.~Golyk, N.~Graham, M.~Kr{\"{u}}ger, and H.~Reid.
  This work is supported by the U.S. Department of Energy under cooperative research agreement Contract Number DE-FG02-05ER41360 (MFM) and the National Science Foundation under Grants No. DMR12-06323 (MK) and NSF PHY11-25915 (RG and MK).

\appendix
\section{Scattering matrices}\label{App: 1}
In this Appendix, we derive the scattering matrix for specific geometries. The Dirichlet boundary conditions are assumed throughout this section.
\subsection{Plate}\label{App: Plate}
  The perturbative scheme introduced in Sec. \ref{Sec: point in 1+1} can be generalized to $d$ dimensions. We designate the coordinates spanning the surface by $\bx_\|$ and the normal coordinate by $z$. The incoming and outgoing waves are identified as
  \begin{equation}
    \Phi^{\inn/\out}_{\omega \bk_\|}(t,\bx)=\frac{1}{{\sqrt{k_\perp}}}\exp(-i \omega t+i \bk_\| \cdot\bx_\| \mp  i k_\perp z)\,,
  \end{equation}
  where $\bk_\|$ is a $(d-1)$-component wavevector parallel to the surface of the plate, and $k_\perp(\omega, \bk_\|)=\sqrt{\omega^2/c^2-\bk_\|^2}$. Consider $\Phi_0$ as the solution to the Dirichlet boundary problem for a static mirror,
  \begin{equation}
    \Phi_0(t,\bx)=\Phi^{\inn}_{\omega \bk_\|}(t,\bx)-\Phi^{\out}_{\omega \bk_\|}(t,\bx)\,.
  \end{equation}
  The scattering matrix can be computed perturbatively by organizing the field as $\Phi=\Phi_0+\Phi_1+\cdots$. The Dirichlet boundary condition, $\Phi(t, \bx_\|,z+q(t,\bx_\|))=0$, to the first order is given by
  \begin{equation}\label{Eq: Phi1}
  \Phi_1(t,\bx_\|,0)=-q(t,\bx_\|)\partial_z\Phi_0(t,\bx_\|, 0),
  \end{equation}
  where $q(t,\bx_\|)$ is the boundary displacement as a function of time and position on the surface.
  Given the (time-dependent) value of the field $\Phi_1$ on the boundary as given by Eq.~(\ref{Eq: Phi1}), one can compute the latter field everywhere in the space by using Green's theorem,
  \begin{equation}\label{Eq: Green's theorem}
    \Phi(x)=\int_{\Sigma} \dd\Sigma_\mu \Phi(x')\, \partial^{\mu} G_{\rm D}(x,x')
  \end{equation}
  where $x$ and $x'$ are spacetime coordinates, and $G_{\rm D}$ is a Green's function satisfying Dirichlet boundary conditions on the plate. The integral in the last equation is over a closed surface including $x$ in its interior. The Green's function, $G_{\rm D}$, is, in Fourier space, given by
  \begin{align}\label{Eq: G-D plate}
     G_{\rm D} (\omega,\bk_\|,z,z' )=\frac{i}{2 k_\perp} e^{i k_\perp z_>}(e^{-i k_\perp z_<}-e^{i k_\perp z_<}).
  \end{align}
  Using Eqs.~(\ref{Eq: Green's theorem}) and (\ref{Eq: G-D plate}), one obtains $\Phi_1$, which in turn gives the scattering matrix as
  \begin{equation}
        S_{\omega+\Omega \bk_\|+\bq, \, \omega \bk_\|}=-2i \, {\tilde q(\Omega,\bq)} \sqrt{k_\perp(\omega, \bk_\|) \, k_\perp(\omega+\Omega, \bk_\|+\bq)}\,.
  \end{equation}

  \subsection{Sphere}\label{App: Sphere}
  In spherical coordinates, the normalized incoming and outgoing waves are defined as
  \begin{equation}
        \Phi^{\inn/\out}_{\omega lm}=\sqrt{\frac{|\omega|}{c}} e^{-i\omega t}\, h_l^{(1,2)}\left(\frac{\omega r}{c}\right)Y_{lm}(\theta, \phi).
  \end{equation}
  The Dirichlet boundary condition for a spherical object in motion is
  \begin{align}\label{Eq: BC sphere}
    \Phi(t, R\hat r+\vec q)=0.
  \end{align}
  In this equation, $R$ is the radius of the sphere, $\hat r$ is the unit vector along the radius, and $\vec q$ is the displacement as a function of time and position on the sphere's surface. For simplicity, we assume that the sphere undergoes a linear (but time-dependent) motion. Equation~(\ref{Eq: BC sphere}) can be expanded in powers of $q$. To the first order, we have
  \begin{equation}
    \Phi_1(t, R\hat r)=-q(t) \cos \theta \partial_r \Phi_0(t, r \hat r )|_{r=R},
  \end{equation}
  where the zeroth-order solution is
  \begin{equation}
    \Phi_0(\omega, \bx)=\Phi^{\inn}_{\omega lm}  +S_{l}(\omega)\Phi^{\out}_{\omega lm},
  \end{equation}
  with $S_{l}(\omega)=-\frac{h_l^{(2)}(\omega R/c) }{h_l^{(1)}(\omega R/c) }$ being the scattering matrix of a static sphere. The Green's function satisfying the Dirichlet boundary conditions on the sphere can be written as
  \begin{align}
    G_{\rm D}(\omega, \bx,& \bx')=\frac{i\omega}{2 c}\sum_{lm} \left( h_l^{(2)}(\omega r_</c)  +S_{l}(\omega) h_l^{(1)}(\omega r_</c)\right)  \nonumber \\
    & \times h_l^{(1)}(\omega r_>/c) Y_{lm }(\theta,\phi) Y^*_{lm }(\theta',\phi')\,.
  \end{align}
  Green's theorem can then be applied to compute $\Phi_1$, or equivalently the scattering matrix as
  \begin{align}
    &S_{\omega+\Omega l'm, \, \omega l m}= \frac{2 i  \tilde q(\Omega)}{c} \, d_{ l l'm} \nonumber \\
    &\times \sqrt{(\omega+\Omega)\,\omega} \, F_l\left(\frac{\omega R}{c}\right) F_{l'}\left(\frac{(\omega+\Omega ) R}{c}\right),
  \end{align}
  with $F$ defined as
  \begin{equation}
    F_l(x)=\frac{1}{x \,h_l^{(1)}(x)}\,.
  \end{equation}
  The constants $d_{l'lm}$ are nonzero only for $l'=l\pm 1$,
  \begin{align}
    d_{l+1\, lm}&=\sqrt{\frac{(l+m+1)(l-m+1)}{(2l+1)(2l+3)}}\,\,,\nonumber \\
    d_{l-1\,lm}&= d_{l\,l-1\,m}\,.
  \end{align}

  \subsection{Disk (cylinder in 2d)} \label{App: Disk}
  In polar coordinates, the normalized incoming and outgoing waves are defined as (up to an irrelevant constant)
  \begin{equation}
    \Phi^{\inn/\out}_0(\omega,\bx)=e^{-i\omega t}H^{(1,2)}(\omega r/c) e^{i m\phi}.
  \end{equation}
  The boundary condition for a circular disk in motion is described similarly to Eq.~(\ref{Eq: BC sphere}) with $\vec r$ being a two-dimensional radial vector. Again we make the assumption that the object undergoes a linear (but time-dependent) motion along the $x$-axis. Equation~(\ref{Eq: BC sphere}) can be expanded as
  \begin{equation}\label{Eq: Phi1 Disk}
    \Phi_1(t, R\hat r)=-q(t) \cos \phi \partial_r \Phi_0(t, r \hat r )|_{r=R},
  \end{equation}
  with $\phi$ being the angle from the $x$ axis. The zeroth order solution, $\Phi_0$, is given by
  \begin{equation}
    \Phi_0(\omega, \bx)=\Phi^{\inn}_{\omega m}  +S_{m}(\omega)\Phi^{\out}_{\omega m},  \end{equation}
  where $S_m(\omega)=-\frac{H_m^{(2)}(\omega R/c) }{H_m^{(1)}(\omega R/c) }$ is the scattering matrix of a static disk. The Green's function subject to Dirichlet boundary conditions on the disk is
  \begin{align}
    G_{\rm D}(\omega,\bx,\bx')= \frac{i}{4}\sum_{m} &\left( H_m^{(2)}(\omega r_</c)  +S_{m}(\omega) H_m^{(1)}(\omega r_</c)\right)  \nonumber \\
     &\times H_m^{(1)}(\omega r_>/c) \, e^{im (\phi-\phi')}\,.
  \end{align}
  With the knowledge of the field on the boundary, Eq.~(\ref{Eq: Phi1 Disk}), one can apply Green's theorem to obtain the field elsewhere in space. One then finds the scattering matrix
   \begin{equation}
    S_{\omega+\Omega m\pm 1, \omega \, m}=\frac{2 i \, \tilde q(\Omega) }{\pi R} \, M_{m}\left(\frac{\omega R}{c}\right) M_{m\pm1}\left(\frac{(\omega+\Omega) R}{c}\right)\!,
  \end{equation}
  where $M$ is defined as
  \begin{equation}
    M_m(x)=\frac{1}{H_m^{(1)}(x)}\,.
  \end{equation}

\end{document}